\documentclass[apj]{emulateapj}
\usepackage{graphicx}
\usepackage[space]{grffile}
\usepackage{latexsym}
\usepackage{amsfonts,amsmath,amssymb}
\usepackage{url}
\usepackage[utf8]{inputenc}
\usepackage{hyperref}
\hypersetup{colorlinks=false,pdfborder={0 0 0}}
\usepackage{textcomp}
\usepackage{longtable}
\usepackage{multirow,booktabs}
\usepackage[english]{babel}
\usepackage{natbib}

\shorttitle{Signatures of Star Cluster Formation}
\shortauthors{Kuznetsova et al.}

\begin{document}


\title{Signatures of Star Cluster Formation by Cold Collapse}


\author{Aleksandra Kuznetsova\altaffilmark{1}, Lee Hartmann\altaffilmark{1},
Javier Ballesteros-Paredes\altaffilmark{2}}
\altaffiltext{1}{Department of Astronomy, University of Michigan, 1085 S. University Ave., Ann Arbor, MI 48109}
\altaffiltext{2}{Centro de Radioastronom\'ia y Astrof\'isica, Universidad Nacional Aut\'onoma de M\'exico, \\
            Apdo. Postal 72-3 (Xangari), Morelia, Michoc\'an 58089, M\'exico}
\email{kuza@umich.edu}


\begin{abstract}
Sub-virial gravitational collapse is one mechanism by which
star clusters may form.  Here we investigate whether this mechanism
can be inferred from observations of young clusters.
To address this question, we have computed SPH simulations of
the initial formation and evolution of a dynamically young star cluster
through cold (sub-virial) collapse, starting with an ellipsoidal,
turbulently seeded distribution of gas, and forming sink particles
representing (proto)stars. While the initial density distributions of the
clouds do not have large initial mass concentrations, gravitational
focusing due to the global morphology leads to cluster formation. We use
the resulting structures to extract observable morphological and
kinematic signatures for the case of sub-virial collapse. We find that
the signatures of the initial conditions can be erased rapidly as the 
gas and stars collapse, suggesting that kinematic observations need
to be made either early in cluster formation and/or at larger scales,
away from the growing cluster core.   Our results emphasize that a dynamically young
system is inherently evolving on short timescales, so that it can be
highly misleading to use current-epoch conditions to study aspects such
as star formation rates as a function of local density. Our simulations
serve as a starting point for further studies of collapse including other 
factors such as magnetic fields and stellar feedback.
\end{abstract}


\citestyle{aa}

\section{Introduction}

\par Most stars form in clusters  \citep{Lada_2003}, and our own  Sun  is likely no exception \citep{Adams_2010}. As a building block of star formation, the dynamical state of molecular clouds plays an important role in determining the properties of star clusters. Theoretical studies of young star clusters have generally started with a relatively dense cloud of gas with ``turbulent'' motions that are approximately virial 
\citep[e.g.,][]{Bate_2003,Scally_2005,Tan_2006,Price_2008,Bonnell_2008,Hennebelle_2012,Myers_2014}.  In the case of numerical simulations, this choice of initial conditions is often made for practical reasons, but this begs the question of how the protocluster gas (and stars) formed (with quasi-virial motions) in the first place.  

Generally speaking, there are two main pictures of the formation of star clusters.  The first posits some type of structuring in the molecular cloud which results in massive clumps which initially are supported by roughly virial (and supersonic) motions \citep{Tan_2006,Hennebelle_2012}, while the second invokes gravitational collapse \citep{Hartmann_2007,Elmegreen_2007,Allison_2009,Allison_2010,Allison_2011}.  
Several studies have addressed the issue of cluster evolution after the stars have been born and investigated whether gas removal by stellar feedback is important \citep{Lada_1984,Kroupa_2000,Kroupa_2001,Scally_2005, Kruijssen_2012, Moeckel_2012}, but relatively little attention has been given to the formation of the protocluster gas \citep{Hartmann_2007,Elmegreen_2007,Hennebelle_2012}.

With the advent of multifiber, high-resolution spectrographs, it has become possible to study the stellar kinematics in star-forming regions efficiently  \citep{Furesz_2008,Tobin_2009,Tobin_2015,Foster_2015,Cottaar_2015}.   The populous Orion Nebula Cluster (ONC), with substantial amounts of molecular gas and continuing star formation is of particular interest as it is potentially young enough \citep[e.g.,][]{Da_Rio_2010} to show kinematic signatures of their formation.  \citet{Furesz_2008} and \citet{Tobin_2009} showed that most of the ONC stars that can be studied optically have radial velocities consistent with those of the molecular gas, with a substantial velocity gradient north of the cluster center in contrast to the southern region.  This kinematic behavior led \citet{Tobin_2009} to suggest that ONC gas and stars are still infalling, though others have argued that the ONC is in expansion  \citep{DaRio_2014,Kroupa_2001} or approximate equilibrium \citep{Tan_2006}.

\par Previous numerical work focused on simulating young clusters with applications to the ONC has primarily utilized N-body simulations without including the gravity of the gas. These studies have shown that initial substructure is rapidly smeared out \citep{Scally_2002,Parker_2012,Parker_2014}, mass segregation such as that observed in the ONC \citep{Hillenbrand_1998} occurs rapidly \citep{Allison_2010}, and that attempts to fit the current distribution of stars in the ONC to a specific scenario  yield a host of possible initial conditions \citep{Scally_2005,Allison_2010,Parker_2012,Parker_2014}.  However, the presence of significant masses of gas in very young clusters means that the self-gravity of the gas can be important in determining the gravitational potential and thus whether clusters like the ONC are bound or not \citep{Hillenbrand_1998}.  

\citet{Proszkow_2009} took an initial step toward including the gas along with cluster stars by modeling it with a static potential included in the N-body calculations.  They showed that sub-virial , with $E_G > E_K$, or cold collapse of the stars in an elongated additional gravitational potential - as likely given the filamentary structure of the dense gas in the ONC - observed at an appropriate angle of inclination to the line of sight could explain the observed stellar radial velocities, which show a difference in kinematics between the northern and southern parts of the cluster \citep{Tobin_2009}.  

To test the subvirial collapse (what we will refer to as cold collapse in this paper)  picture further, it is necessary to include the time-dependent gravitational potential of the gas, which should also be collapsing along with the stars. In the large-scale ``toy model'' of the Orion A cloud, \citet{Hartmann_2007} found that an elongated, rotating, cold collapse model could reproduce the observed morphology of the $^{13}$CO gas and formed a dense massive ``protocluster'' at one end of the cloud.  This model, however, had limited resolution and did not include star formation (creation of sink particles).  Also motivated in part by the structure of Orion A, \citet{Bonnell_2008} simulated star (sink) formation in an initially cylindrical cloud of gas given an initially equilibrium velocity field, which dissipated its kinetic energy and collapsed to form clusters; in this case
the evolving gravitational potential of the gas was included.  However, Bonnell et al.\ were focused on the formation and accretion of brown dwarfs and did not focus on the kinematic signatures of the clusters or cloud.  Similarly, \citet{Bate_2012} presented calculations of cluster formation in a collapsing environment with either a barotropic equation of state or including radiative transfer, but focused mostly on the mass function and relative spatial distribution of low- and high-mass sinks.

\par In this paper, we seek to improve upon the work of \citet{Proszkow_2009} by allowing the potential of the gas to evolve dynamically.  Our simulations are qualitatively similar to that of \citet{Bonnell_2008}, but we focus on a search for spatial or kinematic properties that could provide observational tests of the sub-virial or cold-collapse picture. The calculations also provide qualitative insights into the difficulties of testing models of star formation from observations at a single epoch in a dynamically-evolving environment.  These simulations provide a starting point from which more complex models including magnetic fields and stellar feedback can be developed for comparison with observations.

\section{Numerical method}
\par To simulate cold collapse of a finite cloud with sink formation, we used the smoothed-particle-hydrodynamics (SPH) code Gadget2 \citep{Springel_2005}.  The simulations are evolved on the order of a free fall time, $t\mathrm{_{ff}} = 32\pi (G\rho)^{-1/2} = 0.85$ Myr. The simulation is isothermal (T=10K), so dimensions of the cloud, time evolved, and masses can be rescaled with the sound speed.
 \par In Table \ref{tab:runs}, we list the details of initial conditions of runs we have analysed. Based on the idea of the Orion A model in \citet{Hartmann_2007}, we create an initially homogeneous, ellipsoidal triaxial geometry, which can be thought of as a part of a molecular cloud undergoing collapse.  As star clusters form in a relatively dynamic interstellar medium, some imparted angular momentum is likely, so we put the cloud initially in uniform rotation, along the most elongated axis, again following \citet{Hartmann_2007}.  In addition to runs which serve to replicate the morphology and kinematics of Orion, we include run HR22, an isotropic distribution of gas with no rotation, meant to test which characteristics of infall are generalizable to clusters of arbitrary morphology.
\par Velocity fluctuations were created through the use of a random decaying supersonic turbulent field of various phases, like that of \citep{Stone_1998}, and following the prescriptions of 
\citet{Ballesteros-Paredes_2015}, where wave phases scale as $2\pi/L$ for each direction, where $L$ represents the dimensions of a box defined by the principal axes of the initial ellipsoid. 
Supersonic fluctuations shock the gas, but without constant energy injection, shocks dissipate rapidly. In this context, turbulence serves as a way to create primordial inhomogeneities within a globally collapsing object. 
\par  To simulate star formation, we use the sink implementation from \citet{Jappsen_2005}. While gas particles interact with one another through hydrodynamic and gravitational forces, sinks interact with each other and the particle fluid purely through gravity. Sink formation occurs when a parcel of gas with  $\nabla \cdot u < 0$ eaches some threshold critical density within a certain radius. We set the critical density and radius of sink creation to meet SPH resolution requirements where the critical radius is the Jean's length that corresponds to the minimally resolvable mass \citep{Bate_1997}. The accretion radii, $R_{inner}$ and $R_{outer}$ are scaled according to the critical radius. All particles within $R_{inner}$ are automatically accreted, where it represents the ``physical" boundary of the sink. Particles between $R_{inner}$ and $R_{outer}$ are tested to see if they are gravitationally bound to the sink, (i.e. if  $E_{tot} = E_G + E_K < 0$)
\par  Resolution in SPH refers to the minimum resolvable mass, $M_{res} = 2 M_{tot} N_{neigh} / N_{tot}$ \citep{Bate_1997} and depends on particle parameters as opposed to spatial parameters like grid size. The high resolution runs have $M_{res} = 0.05 M_{\odot}$ and the low resolution runs have $M_{res} = 0.15 M_{\odot}$. Lower resolution runs are suitable for investigating general morphological and kinematic properties. Higher resolution runs are used in situations where better statistics are necessary.

\begin{table*}
\caption{Low and high resolution runs}\label{tab:runs}
\begin{center}
\begin{tabular}{ c c c c c c l }
    Run & Mass $\mathrm{ [M_{\odot}] }$ 
    & \# of particles $[\times 10^6]$ 
    & Dimensions $\mathrm{[pc^3]}$ & Mach \# & $t_{end}$ $\mathrm{[t\mathrm{_{ff}}]}$ & Notes \\ 
    \hline
    \hline
    LR & 2320 & 2 & $3\times 2\times 1$ & 8 & 1.1  & Low resolution, ONC like ellipsoid \\ 
    HR & 2320 & 6 &$3\times 2\times 1$ & 8 & 0.9 & High resolution, ONC like ellipsoid \\ 
    LRb & 1160 & 1 & $3\times 2\times 0.5$ & 8 &  1.3 & Low resolution, flattened ONC like ellipsoid \\ 
    LRc & 580 & 0.5 & $3\times 2\times 0.25$ & 8 & 1.8  & Low resolution, flattened ONC like ellipsoid\\ 
    LRd & 580 & 0.5 & $3\times1\times 0.5$ & 2 & 1.3    & Low resolution, lower Mach number, ellipsoid\\
    HR22 & 1000 & 6 & $1\times 1 \times 1$ & 8 & 1.2 & Cubical, no rotation \citep{Ballesteros-Paredes_2015} \\ 
    \hline
\end{tabular}
\end{center}
\end{table*}

\section{Results}

\subsection{Morphology of Orion-like runs}

\par
\begin{figure}[h!]
\begin{center}
\includegraphics[width=\columnwidth]{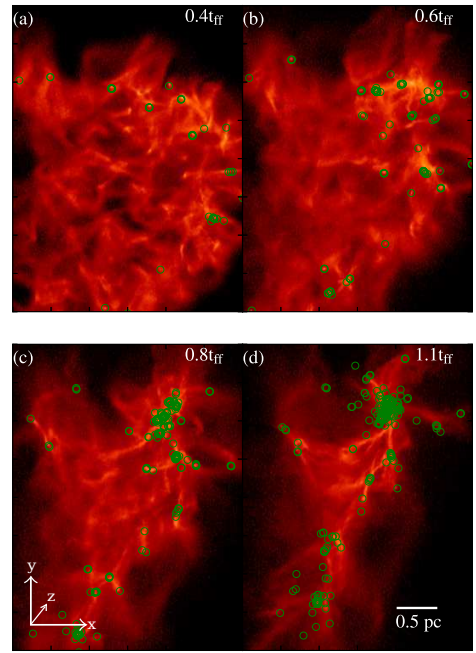}
\caption{\label{fig:prog}
Time series evolution in the XY plane of run LR at (a) 0.4, (b) 0.6 (c) 0.8 and (d) 1.1 $t\mathrm{_{ff}}$. Sink positions are shown by empty circle markers. a) Effect of the initial turbulent velocity field b) Density fluctuations remaining after turbulence dies down, start of vigorous star formation c) Sub clusters start to merge together in area of high potential d) End of simulation with formation of newborn cluster along a dense gaseous filament%
}
\end{center}
\end{figure}

 Figure \ref{fig:prog} shows the time evolution and cluster morphology and star formation over 1.1 $t\mathrm{_{ff}}$ (0.9 Myr) for an Orion-like geometry (run LR) projected in the XY plane. (From this point on, referenced coordinates are based on those marked in bottom left corner of Figure \ref{fig:prog})
Figure \ref{fig:prog}a shows by the end of $0.4 t\mathrm{_{ff}}$ (0.3 Myr), the velocity field shocks, dissipates, and leaves behind density inhomogeneities. Uniform rotation along the longest axis contributes to preferential pile up of material at areas of high curvature, creating regions of high gas potential; star formation is an ongoing process throughout, particularly robust in areas where infall helps coalesce inhomogeneities into flows (seen at $0.6 t\mathrm{_{ff}}$ (0.5 Myr) in Figure \ref{fig:prog}b). At $0.8 t\mathrm{_{ff}}$ (0.7 Myr), the region of high potential has grown deeper as subclusters have started to merge during cluster assembly (Figure \ref{fig:prog}c). Panel d of Figure \ref{fig:prog} is the culmination of $1.1 t\mathrm{_{ff}}$ (0.9 Myr) of evolution; the result is a dense embedded star cluster at one end of a gaseous filament, comparable to the ONC at the end of Orion A.

\par
\begin{figure}[h!]
\begin{center}
\includegraphics[width=\columnwidth]{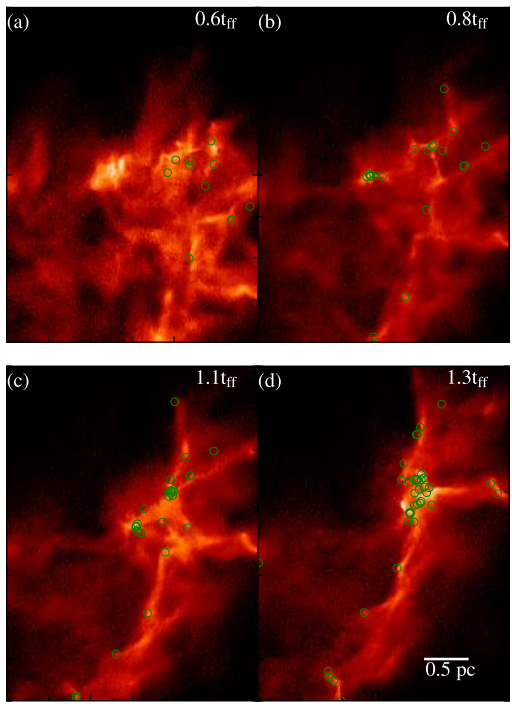}
\caption{\label{fig:lrb} Morphological progression of run LRb over time at (a) 0.6 (b) 0.8 (c)1.1 (d) 1.3 $t\mathrm{_{ff}}$. a) Formation of density inhomogeneities and subfilaments b) Beginning of noticable infall c) Emergence of filamentary structure d) Gas collapses into final filament with knot of condensation at one end, housing clustered stars
}
\end{center}
\end{figure}

 Runs LRb and LRc are intermediaries between the final three-dimensional Orion-like runs, described above, and the limiting case of a rotating sheet as in the toy model of \cite{Hartmann_2007}. As the ellipsoids become more flattened, we can show that the geometry approaches the simplified case seen in HB07 (Figure \ref{fig:lrb}). As such, we demonstrate that the natural outcome of a collapse of a finite rotating object into a filament with a knot of condensation at one end is generalizable to three dimensions.

The cloud and cluster morphology is roughly comparable to that of the simulation of \citet{Bonnell_2008}. They started with a cylindrical molecular cloud 10 pc in length and 3 pc in diameter, with a linear density gradient from one end of the cylinder to the other.  Bonnell et al. also initiated their simulation with a supersonic turbulent velocity field such that the dense end of the cloud was subvirial while the other end was slightly unbound.
The main difference between their results and ours is that they formed two large clusters and one significant group rather than the single main cluster we form, possibly due to a stronger focusing effect in our geometry \citep{Burkert_2004}.

\subsection{Dynamic Evolution}

\subsubsection{Gravity - dynamic driver}

\par As gravity is driving cluster formation, it is useful to characterize the evolution of the gravitational potential over time.
At early times ($t < 0.5 t\mathrm{_{ff}}$), the gas potential is dominant,  determining the location of the future cluster potential minimum.  The increase in densities in this region results in concentrated and star formation centered at
$x = 2.23$, $y = 1.89$, $z = 0.0$; this position does not shift as time proceeds. We use this potential minimum as the cluster center in what follows.

\begin{figure}[h!]
\begin{center}
\includegraphics[width=\columnwidth]{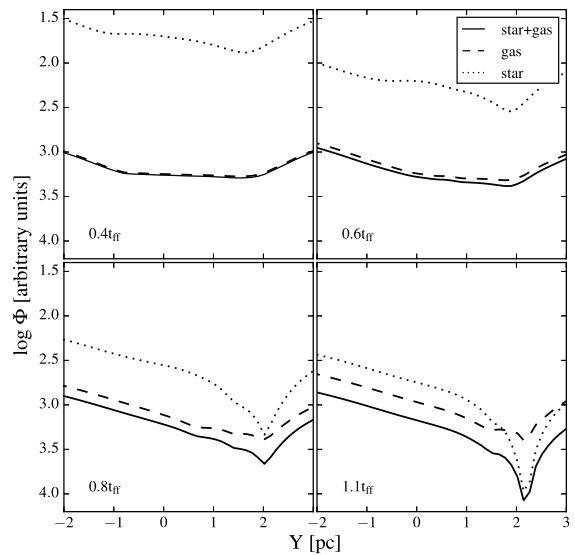}
\caption{\label{fig:pvst} Potential wells sliced along the y axis due to stars(dotted), gas(dashed), and both stars and gas (solid)  over time shown at 0.4, 0.6, 0.8 and 1.1 $\mathrm{t_{ff}}$ shown for run LR%
}
\end{center}
\end{figure}

In Figure \ref{fig:pvst} we show the contributions of stars and gas  to the gravitational potential along a line in the $y$-direction at the cloud midplane ($z = 0$) and at the $x=2.23$ value of the cluster center.As star formation and cluster assembly continues, the stellar potential deepens the existing well on small scales close to the future cluster center. The stellar potential does not begin to be significant until later times ($t> 0.8 t\mathrm{_{ff}}$), when it reaches the same order of magnitude as the potential near the cluster center.

\subsubsection{Kinematic signatures}

\par
\begin{figure}[h!]
\begin{center}
\includegraphics[width=\columnwidth]{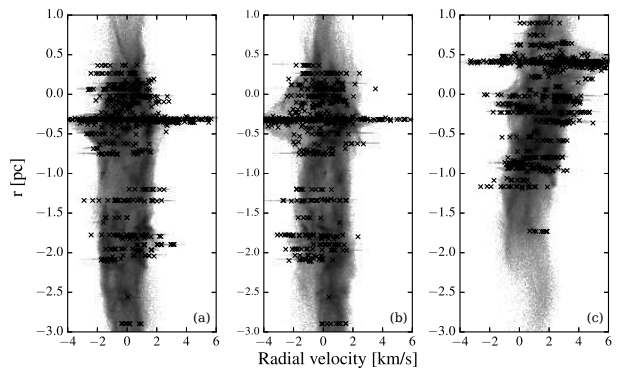}
\caption{\label{fig:pv} Position-radial velocity diagrams for views in (a) the XY plane vs $v_z$ (b) the YZ plane vs $v_x$ (c) the XZ plane vs $v_y$ at $ t= 0.9 \mathrm{ff}$. . 
}
\end{center}
\end{figure}

 \citet{Proszkow_2009} showed that radial velocity gradients as seen in the ONC \citep{Tobin_2009} by a model of cold collapse of stars in a fixed spheroidal gravitational potential of gas, viewed from an appropriate angle. In Figure \ref{fig:pv}, we plot position velocity diagrams for the cluster (run HR) and its surrounding region in various projections (XY, ZY, and XZ planes) at $ t = 0.9 t\mathrm{_{ff}}$ (0.71 Myr) for comparison with the Proszkow et al. results. 

\par
\begin{figure}[h!]
\begin{center}
\includegraphics[width=0.7\columnwidth]{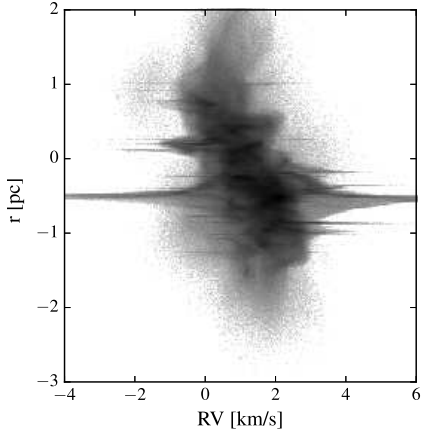}
\caption{\label{fig:pvproj} Position-radial velocity diagrams plotted for a 60 degree projection of the ZY projection about the z axis (panel b of Figure \ref{fig:pv}). Velocity difference between the north and south is 1.0 $\mathrm{km s^{-1}}$
}
\end{center}
\end{figure}

  \citet{Proszkow_2009} found that a gradient such as the one in the ONC was only found when viewing the region at an angle to one of the principal axes of the spheroid. Changing the projection from the ZY plane to the XZ plane by tilting the cluster about 60 degrees about the z axis (tilting from panel b to panel c of Figure \ref{fig:pv})
can reproduce a weak velocity difference of 1.0 $\mathrm{km s^{-1}}$ between the ``north" and ``south" regions of the gas (Figure \ref{fig:pvproj}), smaller than the 2.4 $\mathrm{km s^{-1}}$ difference found in Orion \citep{Tobin_2009}.
This occurs, in part, due to a projection effect achieved by placing regions of the cluster gas and the filament gas in the same plane. To achieve an appreciable gradient, the projection should be in the plane of the initial imparted rotation (i.e. projecting panel a of Figure \ref{fig:pv} toward panel c does not have the same effect). To that effect, runs without added rotation do not develop a significant gradient.

\par
\begin{figure}[h!]
\begin{center}
\includegraphics[width=0.9\columnwidth]{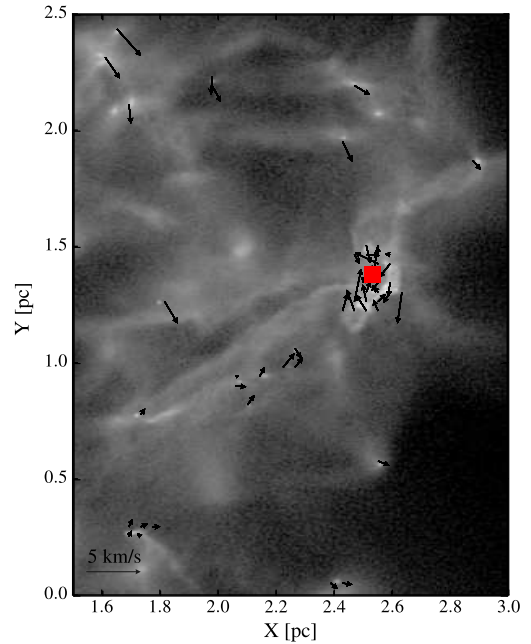}
\caption{\label{fig:gpm} Sample proper motions plotted on the large scale at early times ($t = 0.7 \mathrm{t_{ff}}$) where proper motions are heading toward the current cluster position (denoted by the red square). Background is gas surface density at those locations.}
\end{center}
\end{figure}

 Earlier in the formation process (t = 0.7 $t\mathrm{_{ff}}$, Figure \ref{fig:gpm}), proper motions tend  toward the cluster center. The cluster center at this time is characterized by a large amount of gas and some captured stars. While some proper motions visible in Figure \ref{fig:gpm} are clearly directed toward this location, others are less clear as signatures of infall. Even though we know that infall is occurring, it might be difficult to discern at early times. As the gas potential is the driver of cluster evolution, dense gas structure could play an important part in determining the motions of stars during times before massive cluster assembly.

\par
\begin{figure*}[h!]
\begin{center}
\includegraphics[width=0.8\paperwidth]{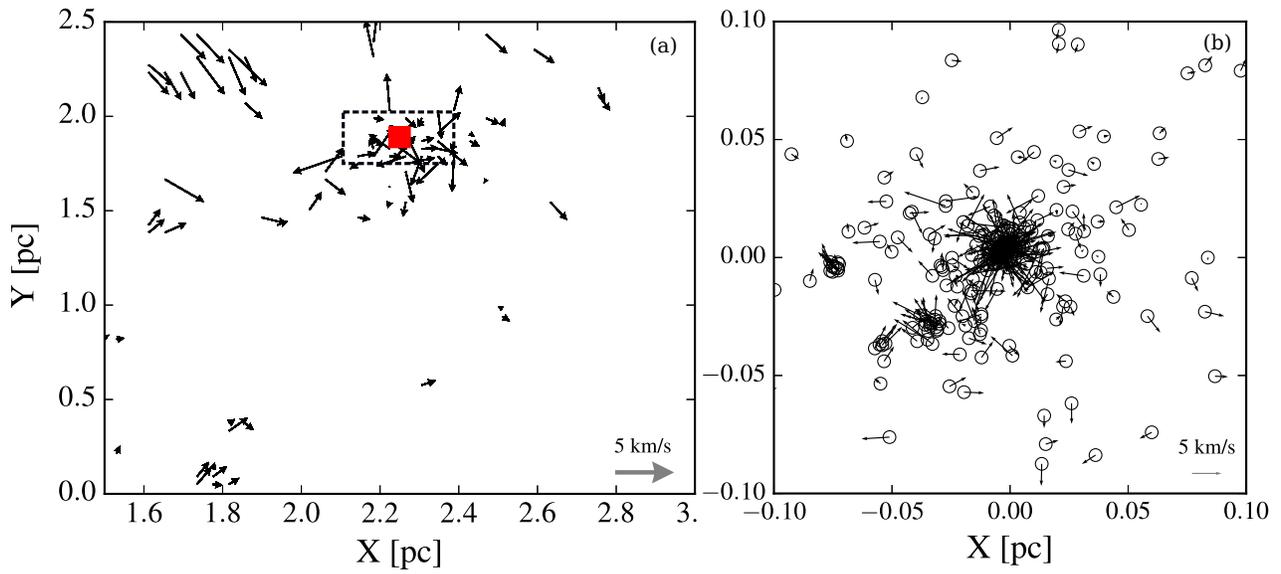}
\caption{\label{fig:localpm} Proper motions of stars at t = 0.9 $\mathrm{t_{ff}}$, post cluster assembly (a) at large scales, proper motions are toward the cluster center (b) enlarged view of the dashed rectangle centered at the cluster center.
}
\end{center}
\end{figure*}

 During the final timestep of the simulation (t = 0.9 $t\mathrm{_{ff}}$), proper motions of sinks far outside the cluster follow the global infall dictated by the gas (Figure \ref{fig:localpm}a), but within the newborn cluster, infall is (unsurprisingly) difficult to distinguish. In Figure \ref{fig:localpm}b, we plot the proper motions of stars within the cluster at the same timestep as Figure \ref{fig:gpm}b,  shown oriented around the cluster center. Using Figure \ref{fig:localpm}, it is evident that infall signatures exist primarily at larger scales, several parsecs outside the cluster. Many of the proper motion vectors do not point directly at the cluster center, due in part to the global angular momentum imparted to the cloud as an initial condition.  However, there is a general,
though weak, signal of collapse in that the
proper motions shown in the upper half of Figure
\ref{fig:gpm}, while the proper motions in the
lower half are much smaller and oriented much
differently (see also Figure \ref{fig:localpm}).

There is little evidence of infall when looking at proper motions within the cluster bounds
(Figure \ref{fig:localpm}, right panel). We can attribute this effect to something like  violent relaxation during the collapse where the rapid growth of the stellar potential in the cluster location can induce a rapid relaxation for particles within the cluster bounds. Two body relaxation is not a plausible mechanism to explain the lack of infall signatures as the timescale for a cluster with the parameters matching our model cluster is on the order of a few Myr, much longer than the freefall time and cluster assembly time for the cluster. Due to dynamical processing associated with global collapse, it appears that there exists a time dependent length scale on which infall signatures manifest.

\par At the end of the simulations, the velocity dispersions of stars within the cluster are nearly isotropic, with values of $3-4 \mathrm{km s^{-1}}$, comparable to that observed in the ONC, while the velocity dispersions of the gas of  $1-2 \mathrm{km s^{-1}}$ are somewhat smaller than ONC values \citep{Tobin_2009,Bally_1987}.  

Decoupling of the stars from gas increases over time as the growth of the stellar dispersion accelerates due to infall. The velocity distributions of the gas do not change significantly over time, unlike the stellar distribution which widens dramatically over a few fractions of the free-fall time. This effect can be attributable to the infall, where the kinematics of the sink particles in the simulation are much more susceptible to the effects of gravity than the gas particles which can experience damping due to the hydrodynamics.

\par To more generally investigate the effects of infall on cluster kinematics, we perform the same analysis on run HR22, an initial setup without elongation or rotation. Similar decoupling between stellar and gas velocities as stellar dispersions grow is observed over the formation and evolution. 

\subsection{Substructure}
\par 
Larger scale infall is responsible for creating the main potential well in which the final cluster forms, while subclusters are created out of the smaller scale primordial density perturbations arising from dissipated shocks made by the initial velocity perturbations. The substructure is initially subvirial, 
but approaches a virial state as the cluster assembles. We can track cluster assembly by looking at the evolution of substructure over time, from a heavily substructured filamentary gas cloud to a centrally concentrated star cluster.
\par Various methods have been used to quantify substructure in a star cluster; the Q parameter  \citep{Parker_2012,Cartwright_2004}, the two point correlation function\citep[][TPCF]{Bastian_2008} and its close relative, the mean surface density of companions (MSDC) \citep{Bate_1998,Larson_1995}. Here we use the TPCF as a proxy for substructure evolution and to track when the cluster begins to assemble into its final distribution. As we are interested in when our final cluster is a smooth power law distribution and do not need to look at multi-scale substructure, the TPCF is an adequate measure of substructure evolution.
\par
\begin{figure}[h!]
\begin{center}
\includegraphics[width=\columnwidth]{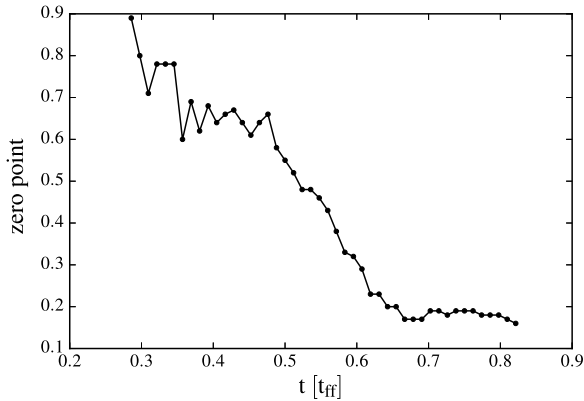}
\caption{\label{fig:zpoint} Zero point of the TPCF shown as a function of time. As structure gets wiped out, the zero point decreases and levels off. 
}
\end{center}
\end{figure}

 We generate a  distribution of all possible distances between sink particles, N, over the cluster area, up to 1.5 pc from the cluster center, and compare it to a random uniform distribution, $N\mathrm{_{ref}}$, scaled over the same area. We define the TPCF as $N/N\mathrm{_{ref}}-1$ and compute it for various times. When the two point correlation function for a certain timestep reaches zero, the sample and reference distributions become the same. Using the zero point of the TPCF at different timesteps, we quantify the changing scale of structure over time. The zero point denotes at what scale (in pc) the distribution and its corresponding reference begin to match up, or, in this case, what the largest scale is at which non uniform structure exists. In Figure \ref{fig:zpoint}, the zero point steadily decreases with time, levelling off at the start of cluster assembly, so that structure becomes relegated to smaller and smaller scales as the cluster forms and homogenizes.

\par
\begin{figure}[h!]
\begin{center}
\includegraphics[width=\columnwidth]{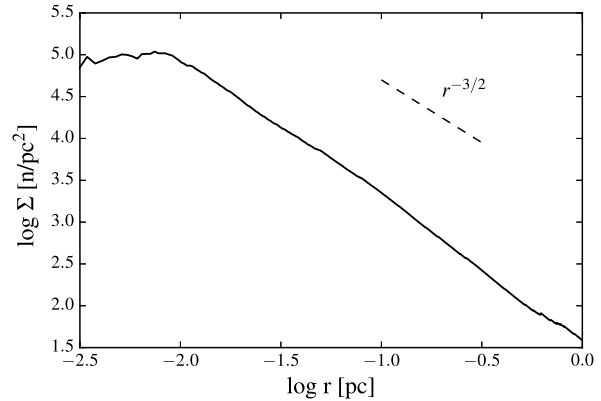}
\caption{\label{fig:starprof} Profile of star surface density in the ONC-like cluster within 1 pc for run HR. Dashed line indicates a profile of $\Sigma \propto r^{-3/2}$. %
}
\end{center}
\end{figure}

 
We find that the final distribution of stars in the newborn cluster is relatively smooth.  The stellar surface density of the local cluster area is comprised of a small inner core decreasing with a power law profile, $\Sigma \propto r^{-3/2}$ (Figure \ref{fig:starprof}) consistent with results from other studies that create profiles with $\Sigma \propto r^{\alpha}$, $\alpha < -1$ \citep{Scally_2002} and observations of the ONC \citep{Hillenbrand_1998, Tan_2006}.
We find a similarly centrally cored surface density profile for our non-ONC like run, HR22, but with a sharper power law drop-off, $\Sigma \propto r^{-2}$. This difference is likely a product of the different geometries between non-ONC and ONC like runs. Run HR22 experienced a more symmetrical collapse without the influence of rotation or an elongated geometry; its initial geometry is probably more efficient at assembling a cluster as it doesn't contain a filament, making the stellar and gas potential wells deeper at the location of the cluster. In cold collapse scenarios, initial geometry is likely an important factor in determining the final cluster geometry.

\subsection{The mass function and mass segregation}
\par
\begin{figure}[h!]
\begin{center}
\includegraphics[width=1\columnwidth]{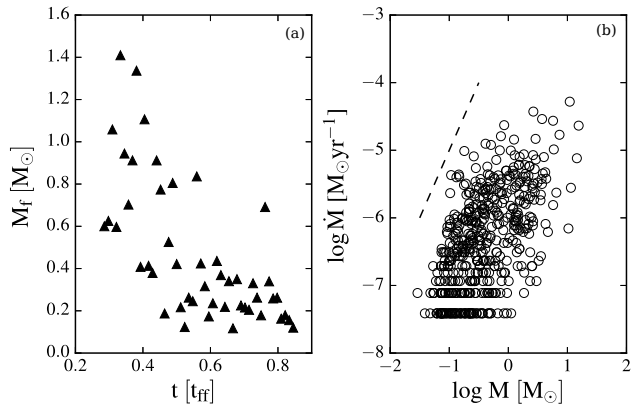}
\caption{\label{fig:tform} (a) Plotted at each time is the average final mass (at $t = 0.9 \mathrm{t_{ff}}$) of the sinks formed during that timestep. (b) Accretion rate for various mass sinks at 0.9 $t\mathrm{_{ff}}$ (run HR).  Discrete accretion rates at low values are a result of minimum mass limits for gas particles in the simulation. Dashed line shown has a slope of 2. 
}
\end{center}
\end{figure}

\begin{figure}[h!]
\begin{center}
\includegraphics[width=\columnwidth]{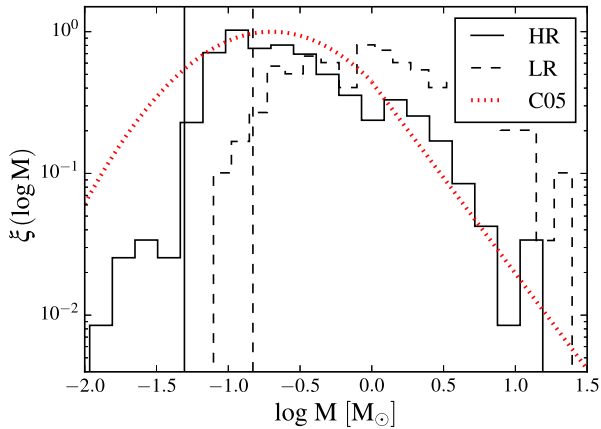}
\caption{\label{fig:IMF} Initial mass functions for high resolution (HR-solid) and low resolution (LR-dashed) runs.Plotted against the IMF from \cite{Chabrier_2005}, normalized uniformly to match numerical IMFs.  Locations of the limits of mass resolution are marked by vertical lines. The turnover for higher resolution runs is well defined and develops earlier due to better statistics. %
}
\end{center}
\end{figure}

 As Figure \ref{fig:prog} shows, sinks form first in 
subclusters along dense filaments and areas of initial gas pileup. Figure \ref{fig:tform}a shows that the (ultimately) more massive sinks form earliest, as we also found in \citet{Ballesteros-Paredes_2015}.
Figure \ref{fig:tform}b shows that on average, larger sinks tend to be more effective accretors. As a result, older sinks have had more time to accrete and gain gravitational influence in an environment becoming more favorable for accretion with time. Previous studies \citep[e.g.][]{Bonnell_2001,Bonnell_2008, Ballesteros-Paredes_2015}, have found that competitive accretion can reproduce a realistic initial mass function, as we see in our own case for high resolution runs with enough dynamic range (Figure \ref{fig:IMF}). 

\par The effects of resolution in this study are seen primarily in the mass function. Figure \ref{fig:IMF} shows the IMF for the low resolution and high resolution runs at the same timestep with vertical lines indicating the minimum resolvable mass for each resolution.Plotted against the empirical model from \cite{Chabrier_2005}, the high resolution run has a well developed power law tail, however, the location of turnover is an artifact of our choice of mass resolution. With a smaller minimum resolvable mass, the higher resolution runs have a better developed turnover in the IMF shifted toward smaller masses. This leads to more readily produced sinks and overall better statistics. The overall morphology and general kinematics remain unchanged between higher and lower resolutions.

\par
\begin{figure}[h!]
\begin{center}
\includegraphics[width=1\columnwidth]{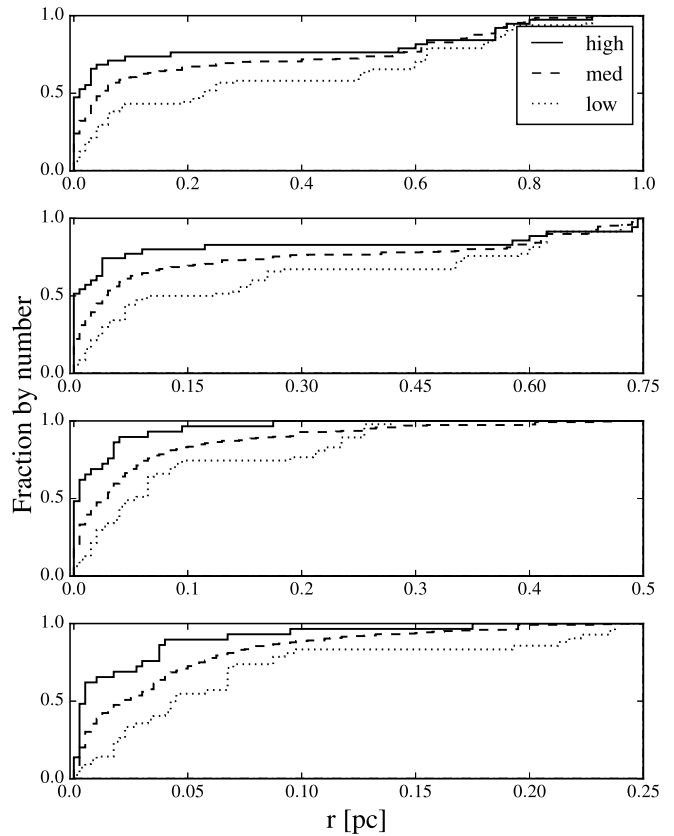}
\caption{\label{fig:seg} Cumulative distributions of stellar populations within (from top to bottom) 1.0, 0.75, 0.5, and 0.25 parsecs from the cluster center. There is clear mass segregation as high mass stars ($M > 2 M_{\odot}$) dominate close to the cluster center. Further mass segregation occurs between intermediate ($0.1 < M < 2 M_{\odot}$) and low ($M < 0.1 M_{\odot}$) mass cluster members as well. (run HR)}
\end{center}
\end{figure}

  Measurements of Orion's population in \citet{Hillenbrand_1998} found clear evidence of mass segregation of the highest mass population, with some evidence for segregation down to lower masses like $1-2 M_{\odot}$. Using the IMF, we divide sinks in the simulation by mass based on different regions of the mass function. Low mass sinks ($M<0.1 M_{\odot}$) are those smaller in mass than the peak of the mass function. High mass sinks ($M>2M_{\odot}$) are those definitively in power law tail of the mass function, with intermediate mass sinks in the region in between. Cumulatively binning the different mass populations in Figure \ref{fig:seg} on different scales, we find evidence for mass segregation between all three populations.

\section{Discussion}

\subsection{Effects of an evolving gas potential}
\par Comparing our results to those of \cite{Proszkow_2009}, we find some similarities in the types of signatures we can detect in the ONC-like cluster - specifically, that the final cluster is fairly elongated and there exists a radial velocity gradient in the gas between the north and south regions of the cloud. While we see a gradient in Figures \ref{fig:pv} and \ref{fig:pvproj}, it is due to projection effects that place the filament and the cluster in the same plane combined with the added initial rotation imparted to the cluster. This emphasizes the importance of projection effects, in agreement with the findings of Proszkow et al. However the importance of the initial rotation component in
our simulation can not be discounted in creating these signatures.
\par The geometry of the gas potential is very similar between our study and that of \citet{Proszkow_2009}, so significant differences in kinematic signatures can be attributed to the presence of an evolving gas potential and the addition of rotation. The extra dynamical processing of the gas that we have added by introducing a non-static gas potential likely ensures that signatures seen in the static case are short-lived in an evolving potential. In addition, the introduction of small scale fluctuations in the gas imparted by the initial supersonic velocity field, can create additional departures from the static case. However, these departures are probably minor due to the fact that the overall gas potential (which dominates the total) is still relatively smooth. 
\par  In a system dominated by infall, we expect to see stellar proper motions pointed toward the center of the cluster. However, as we see  in Figures \ref{fig:gpm} and \ref{fig:localpm}, even though the simulated cluster is relatively young, obvious signatures of infall no longer exist within bounds of the cluster, and few can be seen on the outskirts of the cluster. 
\par  Stellar dispersions grow suddenly and quickly, starting more than halfway through the simulation and growing over a few tenths of a free-fall time.The difference between gas and stellar velocity dispersions is consistent with the study of  
\citet{Foster_2015}, who observed that dense cores in NGC 1333 have smaller velocity dispersions than that of the fully formed stars, suggesting that the growth of stellar velocity dispersions over time occurs as stars become dynamically hotter during infall. They concluded this discrepancy between dense gas and stellar dispersions can be explained by global collapse of a gaseous medium of initially subvirial substructure. We find our simulations to be in agreement with the proposed scenario. Similarly, the velocity dispersion of stars in the ONC is larger than that of the associated $^{13}$CO gas \citep{Tobin_2009}. Time dependence established in observations, like those of NGC 1333 is likely to be erased. The relaxation time for a star cluster with parameters of our final cluster is on the order of a few million years, a loose upper bound that does not take into account gas clearing processes.

\subsection{What is a bound cluster?}

\par
\begin{figure}[h!]
\begin{center}
\includegraphics[width=1\columnwidth]{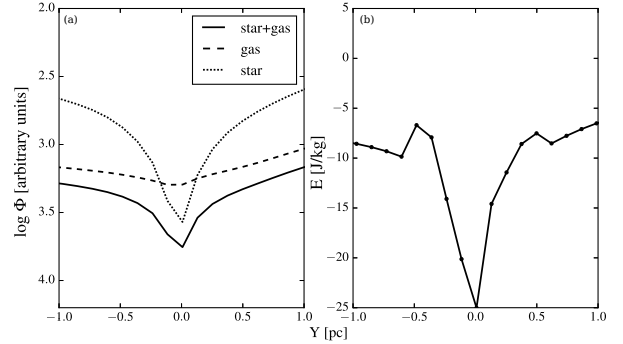}
\caption{\label{fig:eave} Looking at the newborn cluster, we can compare the gravitational potential with the kinetic energy of the stars to look at how bound stars are in regions of the cluster. a) Potential well through the middle of  the cluster site, with gas and star potentials shown separately. b) Average energy (E = KE + PE) per unit mass of stars found in the cluster region, where $E<0$ indicates stars in the region are bound on average.}
\end{center}
\end{figure}

 In Figure \ref{fig:eave}a, we plot a slice of potential through the center of the newborn cluster for run HR (at $t = 0.9 t\mathrm{_{ff}}$), distinguishing between stellar and gas potential. 
The stellar potential is responsible for generating the well for a fraction of the cluster region; the gas potential not only deepens the well, but extends its bounds. Figure \ref{fig:eave}b shows the average energy of stars centered around the newborn cluster at 0.9 $t\mathrm{_{ff}}$ where stars are firmly bound with $E<0$.
Since the distribution is highly centrally concentrated, most of the stars are within 0.3 pc of the center in the deepest part of the potential well shown in Figure \ref{fig:eave}a. 

\par The virial parameter, $\alpha\mathrm{_{vir}} = 5\sigma^2 R/GM$,  is often used to discuss whether a cluster is bound or not . In this study, with full knowledge of the potential, we find the system to be subvirial during formation, only approaching virialized at the end of free fall. This, however, does not ensure the accuracy of the virial parameter. Observational effects and survey design preclude the complete knowledge of the dynamical elements in a system so that without a complete census of the gas in the system, the virial parameter is likely to be misapplied.

\par
\begin{figure}[h!]
\begin{center}
\includegraphics[width=1\columnwidth]{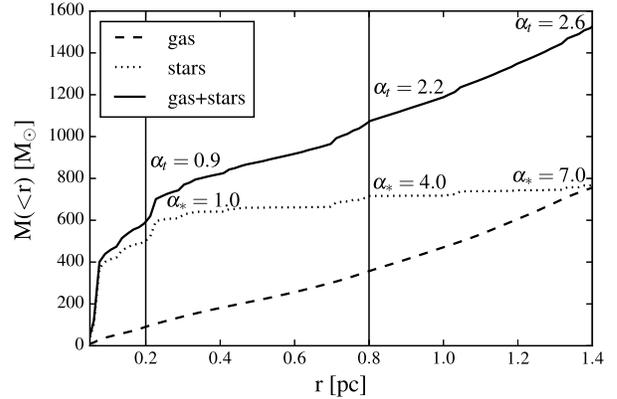}
\caption{\label{fig:massenc} Mass enclosed as a function of radius from the cluster center. Labeled at 0.2, 0.8 and 1.4 pc are virial parameters, $\alpha_*$ and $\alpha_t$ at those radii. $\alpha_*$ - virial parameter comparing kinetic energy in stars and only the stellar potential. $\alpha_t$ - virial parameter comparing the kinetic energy in stars with the total potential.}
\end{center}
\end{figure}

 We find that the virial parameter is highly sensitive to observational techniques. Shown in Figure \ref{fig:massenc}, the enclosed stellar mass levels off within 0.3 pc, but gas mass grows linearly with radius. Labeled at intervals is the calculated virial parameter using only the stellar potential, $\alpha_*$ and the calculated virial parameter using the total potential, including the contribution of the gas mass, $\alpha_t$. We see that $\alpha_*$ is roughly virial when taken within the cluster, but supervirial at and beyond the cluster radius. The virial parameter that takes into account the gas potential,  but not the kinematic properties of the gas, $\alpha_t$, is subvirial within the cluster bounds and virial to supervirial at and outside the cluster radius. When calculating a virial parameter that incorporates all the information at our disposal, kinetic and potential energies for both the gas and the stars, we get that the system is subvirial within the cluster and virial outside the cluster bounds. Since observations of star clusters depend on estimates of cluster size and gas content, using a virial parameter like $\alpha_*$ is unlikely to present an accurate picture of a cluster's dynamical state. 

\subsection{Star forming environments}

\par
\begin{figure}[h!]
\begin{center}
\includegraphics[width=1\columnwidth]{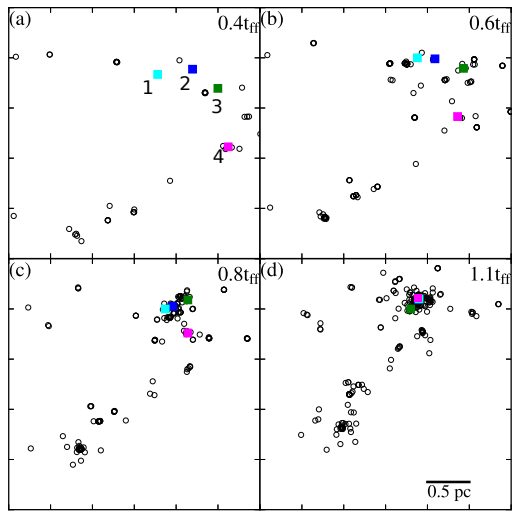}
\caption{\label{fig:massseg} Locations of 4 groups (colored squares) of the 10 most massive sinks in the simulations, numbered to correlate with the groups in Table \ref{tab:groups}. a) Sinks form within subclusters that can travel as groups in areas of high curvature spurred on by infall b) Infall starts to bring massive sinks together c) The  more massive sinks and their subclusters coalesce to make a newborn cluster d) Newborn cluster houses predominantly the most massive sinks in the center.%
}
\end{center}
\end{figure}

 During cold collapse, the environment in the forming cluster/cloud are likely to vary in time and space. Inhomogeneities and filamentary flow create density variations on a variety of spatial scales, while global collapse inherently changes the density of the medium. In assembling the final cluster, stars in this simulation can travel great distances from their birthplaces. We can see in Figure \ref{fig:massseg} and Table \ref{tab:groups} that some of the most massive stars travel several parsecs to get to their final position within the cluster. These groups are born in separate locales, coming from different environments, and accreting material of various density along the way. 

\par
\begin{figure}[t]
\begin{center}
\includegraphics[width=1\columnwidth]{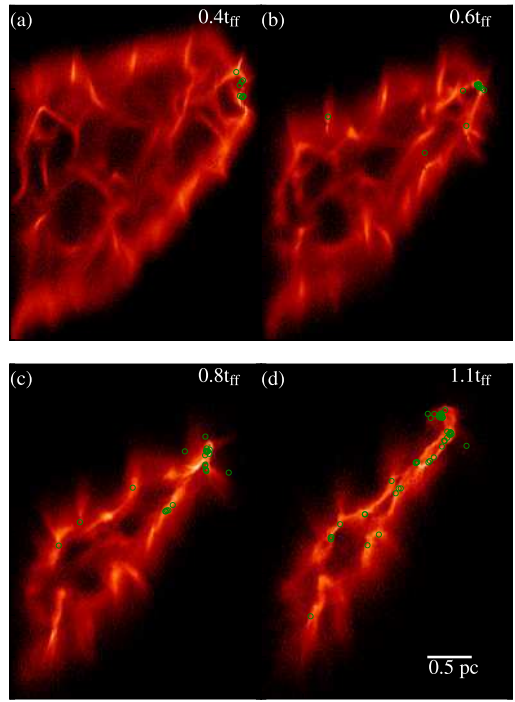}
\caption{\label{fig:fil} Time progression of cluster evolution of simulation where the rotational component is more dominant as the magnitude of initial velocity perturbations has been decreased.
}
\end{center}
\end{figure}

 The substantial migration of accreting sinks (stars) is a result of our initial conditions.  The initial supersonic velocity fluctuations, although rapidly damped, provided non-linear density perturbations in a distributed environment which then collapsed.  Alternatively, one might have started with smaller density perturbations, and arranged collapse to a dense, more filamentary structure in which the sinks would form.  In the latter picture, potentially sinks would not travel far from their birthsites, as seen in many non-clustered regions such as Taurus \citep{Hartmann_2002}.  The spatial and kinematic relations between local gas and stars might thus provide clues to the level of turbulent structure in the initial cloud. As an example, for run LRd where we've lowered the magnitude of all initial velocity perturbations by a factor of 4, we see less clustering and more filamentary star formation(Figure \ref{fig:fil}). The rotation component remains dominant, but star formation occurs along collapsing filamentary structures within the cloud.



\begin{table}[h!]
\caption{Cumulative distances traveled by massive star groups in Figure \ref{fig:massseg}}\label{tab:groups}
\begin{center}
\begin{tabular}{|c | c c c| }
    \hline
    Group & \multicolumn{3}{  c| }{ Distance traveled [pc] at time } \\
     \# & 0.6 t$\mathrm{_{ff}}$ &  0.8 t$\mathrm{_{ff}}$ & 1.1 t$\mathrm{_{ff}}$  \\
    \hline
    \hline
    1 & 0.34 & 0.35 & 0.65\\
    2 & 0.23 & 0.38 & 0.54\\
    3 & 0.41 & 0.90 & 1.28\\
    4 & 0.66 & 1.38 & 2.10\\    
    \hline
\end{tabular}
\end{center}
\end{table}

\par There is evidence for mass segregation in the final cluster, shown in Figure \ref{fig:seg}. However, delineating between a primordial or dynamical origin is difficult as we have shown that stars in a cluster do not have to form anywhere near their final location and can come from different subcluster environments all together.
It is possible that because sinks that become the most massive tend to form first they can thus easily become the first real cluster members during assembly while smaller mass stars form closer to the cluster environment as infall condenses the gas. More than likely, mass segregation is a mix of primordial advantages and dynamical processing.  Our results show that mass segregation is likely due to the geometrical advantages inherent in the set up. However, it is a look at early mass segregation and later effects of feedback processes that disperse gas could very well change the segregation profile of the cluster as seen in studies of feedback in clusters by  \cite{PDE_2015}.

\par
\begin{figure}[h!]
\begin{center}
\includegraphics[width=1\columnwidth]{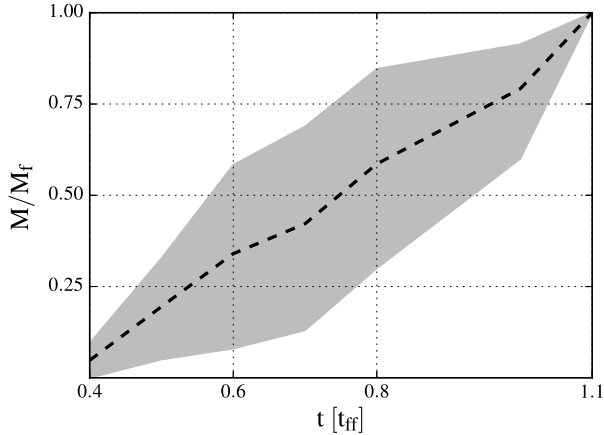}
\caption{\label{fig:history} Line shows the fraction of mass accreted between the snapshots for a typical sink in Figure \ref{fig:massseg}. Shaded region designates upper and lower bounds for sinks in the simulation. Generally, a sink will be over half its final mass by the time it enters the cluster (around t = 0.8 $t\mathrm{_{ff}}$ for the low resolution run corresponding to panel c in Figure \ref{fig:massseg}).
}
\end{center}
\end{figure}

 Another result of the changing landscape that stars form in is that a star forming environment does not simply refer to what is near the cluster, but can refer to varied material that the star forms in and accretes along the way. In fact, we find that, on average, sinks can reach more than half their final mass before even entering the near environs of the cluster (Figure \ref{fig:history}).

\par As a result, measuring star formation efficiency by scaling with local freefall time is misleading as what is meant by ``local" is constantly changing for stars traveling many parsecs through different environments on their way to the final cluster. In addition, an evolving gas potential guarantees that the environment in which stars form and accrete can be very different from the cluster that they are observed in. As brought up in \citet{Allison_2010}, star clusters that we observe only serve as snapshots in a larger and ever shifting story. Attempting to piece together the formation of stars based on the current environment, without thought as to their potential initial conditions, is not likely to be representative of their formation histories.

\par Around 20-40\% of the initial gas mass gets converted to stars at the final stage of the simulation. Using our initial freefall time, the star formation efficiency per freefall time, SFR$\mathrm{_{ff}}$ (defined in \citet{Krumholz_2007}), ranges from 0.2 to 0.36. However, the final cluster environment is, on average, much denser than the inital density. The local free fall time is 0.2 Myr, at most, for near cluster environs. Using this value, the SFR$\mathrm{_{ff}} \lesssim 0.05$, more congruent with estimates made by \citet{Krumholz_2007}. As cold collapse will increase density over time, using current conditions makes metrics like the SFR$\mathrm{_{ff}}$ inherently flawed. 

\par Low star formation efficiency measurements have been used in studies \citep[e.g.]{Krumholz_2007, Tan_2006} to justify initially virialized conditions similar to current cloud conditions. To slow down star formation, mechanisms such as magnetic support and driven turbulence have been invoked. The origin of consistent turbulent driving is still murky, but simulations which inject energy into a system with feedback or turbulence must include gravity, as well, as its effects are demonstrably important.

\section{Summary}

\par Cold collapse has managed to reproduce general morphological and kinematic features of real life clusters. While neglecting ``the kitchen sink", we demonstrate a proof of concept where much of the cluster formation process can be attributed to gravity.  We show that large-scale effects of gravity can naturally produce clusters, even if the near environs of the cluster are dominated by the stars' gravity. 

\par Subvirial initial conditions and an evolving gas potential do not generate long lived kinematic signatures. To be able to confidently identify infall signatures in an observed cluster one would have to catch the cluster early in its formation process and expand the search outside what one might regularly call its bounds. Outside of ideal conditions, it is difficult to adequately confirm what the initial state of a typical star cluster might be. 

\par Future kinematic studies of nearby open clusters will emerge in the coming years, through radial velocity surveys and large scale efforts like GAIA. With future results used in conjunction with a testable model of cold collapse we can make better predictions of the physics involved in a star forming region. While our simulations do not include all potentially important processes (radiative transfer and feedback, magnetic fields), they serve as a starting point for further studies to investigate effects of those processes on the cold collapse paradigm. 

\section{Acknowledgments}
We acknowledge a helpful report from an
anonymous referee.
JBP and LH acknowledge UNAM-PAPIIT grant number IN103012. Numerical simulations were performed either at Miztli, a 4096 core cluster at DGTIC-UNAM, as well as at University of Michigan HPC center, partially supported by UNAM-PAPIIT grant number IN103012 to JBP. AK would like to acknowledge funding by the Rackham Graduate School at the University of Michigan.

\bibliography{biblio}

\end{document}